\documentclass[10pt]{article}
\usepackage{a4wide}
\usepackage[english]{babel}
\usepackage[latin1]{inputenc}
\usepackage{amsmath}
\usepackage{amssymb}
\usepackage[Gray,squaren,thinqspace,thinspace]{SIunits}
\usepackage{graphicx}
\usepackage[small,bf]{caption}

\DeclareMathOperator{\tr}{tr}

\begin{document}

\title{Resolving the instability of the Savvidy vacuum by dynamical gluon mass}
\author{\textbf{David Vercauteren}\thanks{E-mail: David.Vercauteren@UGent.be}, \textbf{Henri Verschelde}\thanks{E-mail: Henri.Verschelde@UGent.be}\\\\
\textit{{\small Ghent University}} \\
\textit{{\small Department of Mathematical Physics and Astronomy}} \\
\textit{{\small Krijgslaan 281-S9, B-9000 Gent, Belgium}}}
\date{}

\maketitle

\abstract{In this paper we apply the formalism of local composite operators as developed by Verschelde \emph{et al.} in combination with a constant chromomagnetic field as considered in the seventies by Savvidy and others. We find that a nonzero $\langle A_\mu^2\rangle$ minimizes the vacuum energy, as in the case with no chromomagnetic field, and that the chromomagnetic field itself is near-to zero. The Nielsen-Olesen instability, caused by the imaginary part in the action, also vanishes. We further investigate the effect of an external chromomagnetic field on the value of $\langle A_\mu^2\rangle$, finding that this condensate is destroyed by sufficiently strong fields. The inverse scenario, where $\langle A_\mu^2\rangle$ is considered as external, results in analogous findings: when this condensate is sufficiently large, the induced chromomagnetic field is lowered to a perturbative value slightly below the applied $\langle A_\mu^2\rangle$.}

\section{Introduction}
In the seventies Savvidy \cite{savvidy} found that the $SU(2)$ Yang-Mills vacuum is unstable against formation of a constant chromomagnetic field. This new vacuum, though, is neither gauge nor Lorentz invariant.

Not much later Nielsen and Olesen \cite{no} showed that the action in this new vacuum has an imaginary part, meaning that the Savvidy vacuum is unstable as well. Ever since then many ways have been explored in order to stabilize this, the most well-known being a dynamical Higgs approach \cite{no, nohiggs} and the "spaghetti vacuum" \cite{noliquidmodel} consisting of a superposition of many domains with different orientations, forming a kind of liquid crystal. Recently new roads of investigation have been explored using the Cho-Faddeev-Niemi decomposition \cite{cfn1,cfn2}, see for example \cite{Cho:2002jt,Kondo:2006ih,Kondo:2004dg}.

In \cite{gsz} Gubarev, Stodolsky and Zakharov proposed that the condensate $\langle A_\mu^2\rangle$ might have a significance in the Yang-Mills vacuum. Although gauge variant, this quantity has been found to be relevant in detecting the condensation of magnetic monopoles in compact QED. It can be shown that this quantity is minimized when working in the Landau gauge. Hence, in that case, $\langle A_\mu^2\rangle_\text{min}$ can be given a gauge invariant interpretation \cite{gsz}.

In \cite{verscheldegn, lcogn} one of us introduced the formalism of local composite operators (LCOs) so as to enable them to calculate this quantity. In \cite{lcos} their method was applied to $SU(N)$ Yang-Mills theory, indeed resulting in a nonzero value for $\langle A_\mu^2\rangle_\text{min}$. A consequence of a nonvanishing value for $\langle A_\mu^2\rangle$ is the dynamical generation of an effective gluon mass.

In this paper we will combine the formalism of LCOs with the constant chromomagnetic background field of Savvidy. In section \ref{LCO} we will give a short review of the LCO formalism in Yang-Mills theory. Section \ref{Seff} will be devoted to calculating the effective action, which will be discussed in section \ref{discussion}. There we search for minima of the action, and we consider the effect of each field on the induced value of the other one. Finally, in section \ref{conclusions} our conclusions will be presented.

\section{LCO formalism} \label{LCO}
In this section we will review the LCO formalism as proposed in \cite{lcos}.

As a first step the gauge is fixed using the Landau condition, i.e. the linear covariant gauge $\partial_\mu A_\mu = 0$ with $\xi \rightarrow 0$. Then, a term
\begin{equation}
\frac12 Z_2 J A_\mu^2
\end{equation}
is added to the Lagrangian density. Here $Z_2$ is a multiplicative renormalization constant and $J$ is the source. As it stands, the theory is not renormalizable. To correct this a new term
\begin{equation}
- \frac12 Z_\zeta\zeta J^2
\end{equation}
has to be added. Here $\zeta$ is a new coupling constant and $Z_\zeta$ is its renormalization factor. This Lagrangian is now multiplicatively renormalizable, as shown in \cite{brst} using a BRST analysis. There are several problems, though.

As a first problem we have introduced a new parameter, $\zeta$, creating a problem of uniqueness. However, it is possible to choose $\zeta$ to be a unique meromorphic function of $g^2$ based on the renormalization group equations. In \cite{lcos} they found using the $\overline{MS}$ scheme in $d=4-\epsilon$ dimensions (up to one-loop order and with $N_c$ the number of colors):
\begin{subequations}
\begin{eqnarray}
\zeta &=& \frac{9}{13}\frac{N_c^2-1}{N_c} \frac1{g^2} + \frac{N_c^2-1}{16\pi^2}\frac{161}{52} \\
Z_\zeta &=& 1 - \frac{g^2N_c}{16\pi^2} \frac{13}{3\epsilon} \\
Z_2 &=& 1 - \frac{N_cg^2}{16\pi^2} \frac3{2\epsilon}
\end{eqnarray}
\end{subequations}

Secondly the presence of the $J^2$ term spoils an energy interpretation for the effective potential defined via the Legendre transform. In order to solve this, a Hubbard-Stratanovich transformation is applied by inserting unity into the path integral:
\begin{equation}
1 = \mathcal N \int[\mathcal D\sigma] \exp-\frac1{2Z_\zeta\zeta} \int \left(\frac\sigma g + \frac12 Z_2A_\mu^2 - Z_\zeta\zeta J\right)^2 d^4x
\end{equation}
with $\mathcal N$ an irrelevant constant. This eliminates the $\frac12 Z_2JA_\mu^2$ and $Z_\zeta\zeta J^2$ terms from the Lagrangian and introduces a new field $\sigma$. The result is:
\begin{equation}
e^{-W(J)} = \int[\mathcal DA_\mu][\mathcal D\sigma] \exp - \int\left(\mathcal L_\text{YM} [A_\mu,c,\bar c] + \mathcal L_\text{LCO} [A_\mu,\sigma] - \frac\sigma g J\right) d^4x
\end{equation}
Herein $\mathcal L_\text{YM}$ is the well-known Yang-Mills Lagrangian with Faddeev-Popov ghosts, fixed in the Landau gauge, and
\begin{equation}
\mathcal L_\text{LCO} [A_\mu,\sigma] = \frac{\sigma^2}{2 g^2 Z_\zeta\zeta} + \frac{1}{2} \frac{Z_2}{g^2 Z_\zeta\zeta} g \sigma A_\mu^a A_\mu^a + \frac{1}{8} \frac{Z_2^2}{Z_\zeta\zeta} (A_\mu^a A_\mu^a)^2
\end{equation}
Now $J$ acts as a linear source for the $\sigma$ field, so that we can straightforwardly compute the effective action $\Gamma(\sigma)$ using the above expressions.

If we compare this to the original expression, we find that the expectation value of $\sigma$ corresponds to the expectation value of the composite operator
\begin{equation}
\sigma = -g\left\langle\frac12 Z_2 A_\mu^2 - Z_\zeta\zeta J\right\rangle
\end{equation}
In the limit $J\rightarrow0$ this operator corresponds (up to a multiplicative factor) to $A_\mu^2$. We can also read off the effective gluon mass in lowest order:
\begin{equation}
m^2 = \frac{N_c}{N_c^2-1}\frac{13}{9} g\sigma
\end{equation}

\section{Effective action} \label{Seff}
\subsection{Introductory matters}
We now proceed to combine the formalism of a constant chromomagnetic background field with the one of massive gluons using the LCO formalism.

Since we are now working with a background field, it is more appropriate to use the Landau background gauge \cite{Abbott:1981ke} $\mathcal D_\mu[\hat A] A_\mu = 0$ instead of the usual Landau gauge prescription $\partial_\mu A_\mu = 0$. Here $\hat A_\mu$ is the background field. In order to do so, some alterations are in order.

A BRST analysis (for BRST in the background gauge, see for example \cite{brstbackground}) shows that, in order for the LCO formalism to stay renormalizable, the condensate $A_\mu^2$ must be replaced by
\begin{equation}
A_\mu^2 - \hat A_\mu^2 = \mathcal A_\mu^2 + 2 \mathcal A_\mu \hat A_\mu
\end{equation}
with $A_\mu$ the total gauge field and $\mathcal A_\mu$ the quantum fluctuations, $A_\mu = \mathcal A_\mu + \hat A_\mu$.

This replacement will change nothing in the expressions for $\zeta$ and the renormalization constants. In the limit $\hat A_\mu = 0$ these must reduce to the original expressions, and since these constants are dimensionless and $\hat A_\mu$ is the only dimensionful parameter they could otherwise depend on, they will not change when switching on a nonzero background field.

As a result, the action we depart from is given by
\begin{multline}
\mathcal{L} = \frac{1}{4}(F_{\mu\nu}^a [\mathcal A_\mu+\hat A_\mu])^2 - \frac1{2\xi} (\mathcal{D}_\mu[\hat A_\mu] \mathcal A_\mu^a)^2 + \bar{c}^a \mathcal{D}_\mu[\hat A_\mu] \mathcal{D}_\mu[\mathcal A_\mu+\hat A_\mu] c^a \\ + \frac{\sigma^2}{2 g^2 Z_\zeta\zeta} + \frac{1}{2} \frac{Z_2}{g^2 Z_\zeta\zeta} g \sigma ((\mathcal A_\mu^a)^2+2\mathcal A_\mu^a\hat A_\mu^a) + \frac{1}{8} \frac{Z_2^2}{Z_\zeta\zeta} ((\mathcal A_\mu^a)^2+2\mathcal A_\mu^a\hat A_\mu^a)^2
\end{multline}

For simplicity, we will work in $SU(2)$. If we choose the background field $B_\mu^a$ to be a chromomagnetic field in the $z$-direction in space and in the 3-direction in isospace, we can write
\begin{equation}
\hat A_\mu^a = Hx_1\delta^{a3}g_{\mu2}
\end{equation}
With this expression, the effective potential at one loop is given by:
\begin{eqnarray}
V_\text{eff} &=& \frac{1}{2}H^2 + \frac{\sigma^2}{2 g^2 Z_\zeta\zeta} - \log\det(\mathcal{D}^2) \\
&& + \frac12 \log\det\left(g_{\mu\nu}\delta^{ab} \frac{Z_2}{g^2 Z_\zeta\zeta} g \sigma - g_{\mu\nu} \mathcal{D}^2_{ab} + \left(1-\frac1\xi\right) (\mathcal{D}_\mu\mathcal{D}_\nu)^{ab} + 2g\epsilon^{ab3} H S^3_{\mu\nu} \right) \nonumber
\end{eqnarray}
where the limit $\xi\rightarrow0$ is implied, and
\begin{equation}
S^3_{\mu\nu} = \begin{pmatrix} 0 &&& \\ & 0 & -1 & \\ & 1 & 0 & \\ &&& 0 \end{pmatrix}
\end{equation}

\subsection{Spectrum of $\mathcal D^2$}
We start by calculating the determinant of the ghost operator.

If we use the eigenbasis of $\epsilon^{ab3}$, we first have the "3" ghosts, for which the covariant derivative reduces to an ordinary one, and then we have the "$+$" and the "$-$" ghosts with eigenvalues $\pm\imath$. For those last ones the covariant derivative equals $\mathcal D_\mu = \partial_\mu \pm \imath gHx_1 g_{\mu2}$.

The "3" ghosts give a trivial contribution of $\tr\log\partial^2 = 0$.

For the "$+$" and the "$-$" ghosts we need the eigenfunctions of $\mathcal D^2 = \partial^2 \pm 2\imath H x_1 g \partial_2 - g^2H^2x_1^2$. This is a harmonic oscillator, and we readily find:
\begin{equation}
-\mathcal D^2 e^{\imath \vec x\cdot\vec p} \psi_n \left(\sqrt{gH}x_1\pm \frac{p_2}{\sqrt{gH}}\right) = (gH(2n+1)+p_3^2+p_4^2)e^{\imath \vec x\cdot\vec p} \psi_n \left(\sqrt{gH}x_1\pm \frac{p_2}{\sqrt{gH}}\right) \label{harmonic}
\end{equation}
where $\psi_n$ ($n\in\mathbb N$) is the $n$th eigenfunction of the harmonic oscillator and with $\vec a = (a_3, a_4)$. We get with dimensional ($d=4-\epsilon$) and zeta function regularization ($\sum_n (n+q)^{-s} = \zeta(s;q)$, the Hurwitz zeta function):
\begin{eqnarray}
\tr\log \mathcal D^2 &=& 2 \frac{gH}{2\pi} \sum_{n=0}^{+\infty} \int \frac{d^{d-2}p}{(2\pi)^{d-2}} \ln\left(gH(2n+1) + \vec p^2 \right) \\
&=& \frac{g^2H^2}{3(4\pi)^2} \left(\frac2\epsilon+1-\ln\frac{2gH}{\bar{\mu}^2}- 12\zeta'(-1) - \ln(2)\right) \label{ghost}
\end{eqnarray}
where $\zeta(s) = \sum_n n^{-s}$ is the Riemann zeta function. Here, the factor of two comes from the contributions of both "$+$" and "$-$" ghosts.

\subsection{Spectrum of the gluons}
The gluons can be split into two classes: the ones obeying the Landau gauge prescription $\mathcal D_\mu \psi_\mu = 0$, giving
\begin{equation}
\frac1 2 \tr_{\mathcal D_\mu \psi_\mu = 0}\log\left(g_{\mu\nu}\delta^{ab} \frac{Z_2}{g^2 Z_\zeta\zeta} g \sigma - g_{\mu\nu} \mathcal{D}^2_{ab} + 2g\epsilon^{ab3} H S^3_{\mu\nu}\right)
\end{equation}
and the ones not satisfying the prescription, giving
\begin{equation}
\frac1 2 \tr_{\mathcal D_\mu \psi_\mu  \not= 0}\log \mathcal{D}_\mu\mathcal{D}_\nu + \text{constant}
\end{equation}
where the irrelevant constant part contains $\lim_{\xi\rightarrow0} \ln \xi$.

The spectrum of this last operator can be reduced to the spectrum of $\mathcal D^2$. If $\psi_\mu$ is an eigenfunction of $\mathcal{D}_\mu\mathcal{D}_\nu$ with eigenvalue $k \not= 0$, we also have that $\mathcal{D}^2 \mathcal{D}_\mu \psi_\mu = k \mathcal{D}_\mu \psi_\mu$ so that $\mathcal D_\mu \psi_\mu$ is an eigenfunction of $\mathcal D^2$ with eigenvalue $k$. This means that all eigenvalues of $\mathcal D_\mu \mathcal D_\nu$ are also eigenvalues of $\mathcal D^2$. Conversely, if  $f$ is an eigenfunction of the operator $\mathcal D^2$ with eigenvalue $p$, then $\mathcal{D}_\mu f$ will be an eigenfunction of $\mathcal D_\mu \mathcal D_\nu$ with the same eigenvalue. Thus we see that these two operators have an identical spectrum and we can write
\begin{equation}
\tr_{\mathcal D_\mu \psi_\mu  \not= 0}\log \mathcal{D}_\mu\mathcal{D}_\nu = \tr\log \mathcal{D}^2
\end{equation}
The expression on the right-hand side has been calculated above. Since the ghosts will come with a factor $-1$ and the gluons with a factor $1/2$, exactly minus one half of the result given there will remain.

For the gluons fulfilling the gauge prescription, we start with the "3" gluons. A straightforward calculation yields for the three polarizations in dimensional regularization:
\begin{equation}
\frac{3-\epsilon}2 \tr \log\left(\frac{Z_2}{g^2 Z_\zeta\zeta} g \sigma - \partial^2\right) = -\frac{3Z_2^2\sigma^2}{4g^2Z_\zeta^2\zeta^2(4\pi)^2} \left(\frac{2}\epsilon + \frac{5}6 - \ln\frac{Z_2\sigma}{2gZ_\zeta\zeta\bar{\mu}^2}\right)
\end{equation}

Secondly there are the "$+$" and the "$-$" gluons:
\begin{equation}
\frac1 2 \tr_{\mathcal{D}_\mu \psi_\mu = 0}\log\left(g_{\mu\nu} \frac{Z_2}{g^2 Z_\zeta\zeta} g \sigma - g_{\mu\nu} \mathcal{D}^2 \pm 2\imath g H S^3_{\mu\nu} \right)
\end{equation}
with $\mathcal D_\mu = \partial_\mu \pm \imath gH x_1g_{\mu2}$. We start with the "$+$" gluons. We now pass to the polarization basis wherein $S^3_{\mu\nu}$ is diagonal. We get:
\begin{align}
S^3_{\mu\nu} &= \begin{pmatrix} \imath &&& \\ & -\imath && \\ && 0 & \\ &&& 0 \end{pmatrix} & \mathcal{D}_\mu &= \begin{pmatrix} \sqrt{gH} \hat a & \imath \sqrt{gH} \hat a^\dagger & \partial_3 & \partial_4 \end{pmatrix}
\end{align}
with $\hat a$ and $\hat a^\dagger$ the ladder operators of the harmonic oscillator from equations \eqref{harmonic}. This reduces the problem to four one-dimensional harmonic oscillators with the same eigenfunctions as in \eqref{harmonic}. The eigenvalues are $gH(2n+1+2s) + p_3^2 + p_4^2 + Z_2\sigma/(gZ_\zeta\zeta)$ with $s=-1,1,0,0$ the spin eigenvalue of the state.

Now we have to restrict the spectrum according to the Landau background gauge. For this purpose we construct the following vector functions from the scalar eigenfunctions
\begin{equation}
f_n = e^{\imath \vec x\cdot\vec p} \begin{pmatrix} e_1 \psi_{n+1} \left(\sqrt{gH}x_1+ \frac{p_2}{\sqrt{gH}}\right) \\ e_2 \psi_{n-1} \left(\sqrt{gH}x_1+ \frac{p_2}{\sqrt{gH}}\right) \\ e_3 \psi_n \left(\sqrt{gH}x_1+ \frac{p_2}{\sqrt{gH}}\right) \\ e_4 \psi_n \left(\sqrt{gH}x_1+ \frac{p_2}{\sqrt{gH}}\right) \end{pmatrix} \quad,\qquad n = -1,0,1,2\ldots
\end{equation}
where $\psi_n$ with $n$ negative is defined to be zero. The vector $e_\mu$ is a polarization vector. These functions have eigenvalues $Z_2\sigma/(g Z_\zeta\zeta) + gH(2n+1) + p_3^2 + p_4^2$. To see whether they obey the gauge condition, we calculate
\begin{equation}
\mathcal{D}_\mu f_n^\mu = \begin{cases} 0 & n = -1 \\ \left(e_1 \sqrt{gH} + \imath e_3 p_3 + \imath  e_4 p_4\right) e^{\imath \vec x\cdot\vec p} \psi_0(\ldots) & n = 0 \\ \left(e_1 \sqrt{gH}\sqrt{n + 1} + \imath e_2 \sqrt{gH} \sqrt{n} + \imath e_3 p_3 + \imath  e_4 p_4 \right) e^{\imath \vec x\cdot\vec p} \psi_n(\ldots) & n > 0 \end{cases}
\end{equation}
We conclude that, for $n=-1$, there is but one polarization with a contribution of $1/2 \tr\log (Z_2\sigma/(g Z_\zeta\zeta) - gH + p_3^2 + p_4^2)$. For $n=0$, of the three, one is eliminated by the gauge prescription, leaving us with 2 polarizations ($2-\epsilon$ in dimensional regularization) each contributing $1/2 \tr\log (Z_2\sigma/(g Z_\zeta\zeta) + gH + p_3^2 + p_4^2)$. For $n>0$ we have the usual 3 ($3-\epsilon$) polarizations with the usual contribution. For ease of calculation, we calculate the second and third groups together with 3 polarizations, so that we have to subtract the contribution of $n=0$ exactly once.

The gluons with $n=-1$ give
\begin{equation}
\frac{gH}{4\pi} \int\frac{d^{2-\epsilon}p}{(2\pi)^{2-\epsilon}} \log \left(\frac{Z_2\sigma}{g Z_\zeta\zeta} - gH + p^2\right) = \frac{gH\left(\frac{Z_2\sigma}{g Z_\zeta\zeta} - gH\right)}{(4\pi)^2} \left(\frac2\epsilon + 1 - \ln\frac{\frac{Z_2\sigma}{g Z_\zeta\zeta} - gH}{\bar{\mu}^2}\right)
\end{equation}
For the gluons with $n=0$ there remains:
\begin{equation}
-\frac{gH}{4\pi} \int\frac{d^{2-\epsilon}p}{(2\pi)^{2-\epsilon}} \log \left(\frac{Z_2\sigma}{g Z_\zeta\zeta} + gH + p^2\right) = -\frac{gH\left(\frac{Z_2\sigma}{g Z_\zeta\zeta} + gH\right)}{(4\pi)^2} \left(\frac2\epsilon + 1 - \ln\frac{\frac{Z_2\sigma}{g Z_\zeta\zeta} + gH}{\bar{\mu}^2}\right)
\end{equation}
And finally all the other states contribute
\begin{multline}
(3-\epsilon)\frac{gH}{4\pi} \sum_{n=0}^{+\infty} \int\frac{d^{2-\epsilon}p}{(2\pi)^{2-\epsilon}} \log \left(\frac{Z_2\sigma}{g Z_\zeta\zeta} + gH(2n+1) + p^2\right) \\
= -\frac{\frac{3Z_2^2\sigma^2}{g^2 Z_\zeta^2\zeta^2}-g^2H^2}{4(4\pi)^2} \left(\frac2\epsilon+\frac13-\ln\frac{2gH}{\bar{\mu}^2}\right) + \frac{6(gH)^2}{(4\pi)^2} \frac{\partial\zeta}{\partial s}\left(-1;\frac12+\frac{Z_2\sigma}{2g^2H Z_\zeta\zeta}\right)
\end{multline}
Here, $\zeta(s;q)$ denotes the analytic continuation of the Hurwitz zeta function, which for first argument greater than one is defined as
\begin{equation}
\zeta(s;q) = \sum_{k=1}^{+\infty} (k+q)^{-s}
\end{equation}
or by its integral representation
\begin{equation}
\zeta(s,q)=\frac{1}{\Gamma(s)} \int_0^\infty \frac{t^{s-1}e^{-qt}}{1-e^{-t}}dt \label{zetaint}
\end{equation}
The derivative $\partial\zeta/\partial s$ stands for the derivative with respect to the first argument. In this last calculation we have made use of the relation between the Hurwitz zeta function and the Bernoulli polynomials, in our case:
\begin{equation}
\zeta(-1,x) = -\frac{B_2(x)}2 = - \frac{x^2}2 + \frac x2 - \frac1{12}
\end{equation}

The "$-$" gluons give exactly the same contribution, so that the above expressions must be multiplied by a factor of two.

\subsection{Total}
\begin{figure}\begin{center}
\includegraphics[width=7cm]{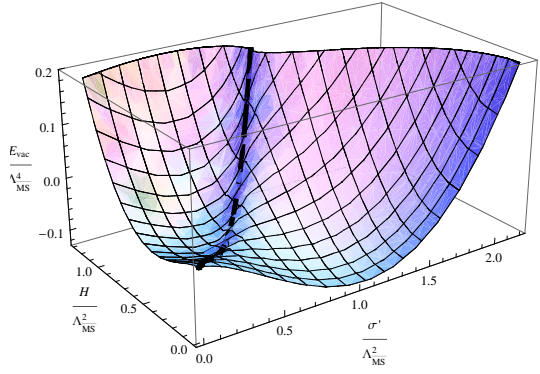}
\caption{The effective action \eqref{Leff} in function of $H$ and $\sigma'$. The black line denotes $\sigma'=H$, where the imaginary part of the action vanishes. \label{Seff3d}}
\end{center}\end{figure}

If we sum all the terms we have calculated, and we substitute the values for the renormalization constants, we get:
\begin{eqnarray} \label{Leff}
V_\text{eff} &=& \frac12H^2 + \frac{27}{26}\frac{\sigma'^2}2 \nonumber \\
&& -\frac{9g^2\sigma'^2}{4(4\pi)^2} \left(\frac12 + \frac{161}{78} - \frac13\ln\frac{g\sigma'}{\bar{\mu}^2} - \frac23\ln\frac{2gH}{\bar{\mu}^2}\right) \nonumber \\
&& - \frac{2g^2H\sigma'}{(4\pi)^2} \ln\frac{\sigma' - H}{\sigma' + H} \nonumber \\
&& - \frac{g^2H^2}{(4\pi)^2} \left(4 - 2 \ln\frac{g\sigma' - gH}{\bar{\mu}^2} - 2 \ln\frac{g\sigma' + gH}{\bar{\mu}^2} + \frac13\ln\frac{2gH}{\bar{\mu}^2} - 2\zeta'(-1) - \frac16\ln(2)\right) \nonumber \\
&& + \frac{12g^2H^2}{(4\pi)^2} \frac{\partial\zeta}{\partial s}\left(-1;\frac12+\frac{\sigma'}{2H}\right)
\end{eqnarray}
where we have set
\begin{equation}
g\sigma' = \frac{26}{27} g\sigma
\end{equation}
so that the effective gluon mass squared is $m_\text{eff}^2 = g\sigma'$. The real part of \eqref{Leff} is plotted in figure \ref{Seff3d}.

In the limit $H \rightarrow 0$ this expression reduces to the one obtained in \cite{lcos}, and when taking $\sigma\rightarrow0$ we get the result of Nielsen and Olesen \cite{no} modulo some differences due to the use of another subtraction scheme and gauge.

When $H>\sigma'$ our potential \eqref{Leff} has an imaginary part
\begin{equation}
-\frac{\imath g^2H}{8\pi} (H-\sigma')
\end{equation}
which reduces to the Nielsen and Olesen result for $\sigma'=0$. It turns out to vanish for $H \leq\sigma'$, so that the Nielsen-Olesen problem of the Savvidy vacuum is then resolved, as predicted would happen by Nielsen and Olesen \cite{no} themselves.

\section{Discussion}\label{discussion}
In the next two subsections, we will find that the minimum of the effective potential is for $H=0$ (or virtually zero) and $\sigma'$ the value calculated in \cite{lcos}.

In order to do so, we will consider two cases: first $H$ will be considered as an external field and $\sigma'$ as an effective gluon mass induced by quantum effects, and next we will investigate the influence of a nonzero $\sigma'$ on the value of the Savvidy field. When looking at small values of the external fields, the analyses can be done analytically by expanding the potential in this small parameter. The scale can then be chosen according to renormalization group considerations. For bigger values of the external fields, however, we proceed numerically. In this last case the scale $\bar\mu^2$ is, for the ease of calculation, fixed equal to $\bar\mu^2 = \unit{4.12}{\Lambda_{\overline{\text{MS}}}^2}$, the value of $g\sigma'$ in the global minimum of the effective action. In that point the coupling constant is reasonably small:
\begin{equation}
\frac{g^2}{8\pi^2} = \frac{36}{187} \approx 0.19
\end{equation}

\subsection{Effect of $H$ on $\sigma'$}
If $H$ is set to zero, the effective potential has a perturbative extremum (a maximum) in $\sigma'_\text{p}=0$ and a non-perturbative minimum at
\begin{equation} \label{sigma}
g\sigma'_\text{np} = \Lambda_{\overline{\text{MS}}}^2 e^{\frac{24\pi^2}{11g^2}} = \unit{4.12}{\Lambda_{\overline{\text{MS}}}^2}
\end{equation}
where the scale was chosen equal to $g\sigma'_\text{np}$.

For small $H$ the equations can be expanded in a series in $H$:
\begin{multline}
V_\text{eff}(H,\sigma') = \frac{27}{26}\frac{\sigma'^2}2 -\frac{9g^2\sigma'^2}{4(4\pi)^2} \left(\frac56 + \frac{161}{78} - \ln\frac{g\sigma'}{\bar{\mu}^2} \right) \\ + \frac12H^2 - \frac{g^2H^2}{(4\pi)^2} \left(\frac12 - \frac72 \ln\frac{g\sigma'}{\bar{\mu}^2} - \frac16\ln\frac{2gH}{\bar{\mu}^2} - 2\zeta'(-1) - \frac16\ln(2) \right) \\ + \mathcal O(H^3\ln H)
\end{multline}
To obtain this, we used the expansion of the Hurwitz zeta function for large arguments, which can be straightforwardly calculated from the integral representation \eqref{zetaint}. From this can easily be obtained that, up to this order,
\begin{equation}
g\sigma'_\text{np} = g\sigma'_{H=0} -\frac{7 gH^2}{9\sigma'_{H=0}}
\end{equation}
so that $\sigma'_\text{np}$ decreases with a raising of $H$. The vacuum energy changes like
\begin{equation}
E_\text{vac} = E_{H=0} - \frac4{13}H^2 + \frac{g^2H^2}{(4\pi)^2} \left(\frac{1231}{156} + \frac16\ln\frac{4H}{\sigma'_{H=0}} + 2\zeta'(-1) \right) + \ldots
\end{equation}

\begin{figure}\begin{center}
\includegraphics[width=7cm]{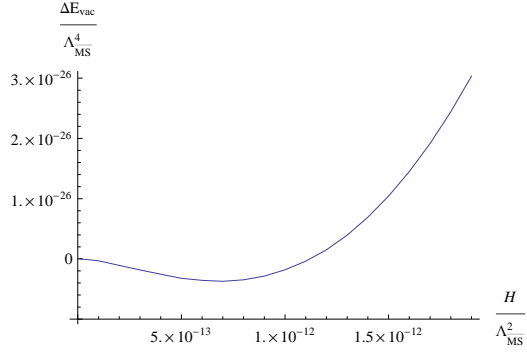}
\caption{The difference between $E_\text{vac} (H)$ and $E_\text{vac} (H=0)$ for very small values of $H$ in the non-perturbative minimum for $\sigma'$. A shallow minimum (order \unit{10^{-27}}{\Lambda_{\overline{\text{MS}}}^2}) is seen for $H \approx \unit{7\times10^{-13}}{\Lambda_{\overline{\text{MS}}}^2}$. \label{kleineH}}
\end{center}\end{figure}
We see that for very small $H$ the term of order $H^2\ln H$ will dominate, lowering the vacuum energy. Very fast, though, this term will be supplanted by the terms of order $H^2$, and the energy will start increasing again. The effective potential in this regime is depicted in Figure \ref{kleineH}. The lowest value is reached when
\begin{equation}
H = \sigma' \exp\left(\frac{384\pi^2}{13g^2}-\frac{1257}{26}-2\ln2-12\zeta'(-1)\right) = \unit{3.92\times10^{-13}}{\sigma'}
\end{equation}
Since this result is astronomically small, there is no reason why it wouldn't disappear when higher-order corrections or any other effects are taken into account. For all practical purposes one can say that the vacuum energy is lowest when $H=0$ and $\sigma'$ has the value given in \eqref{sigma}. One would expect terms containing $\ln H$ to be replaced with $\ln(H+\sigma)$ when switching on a mass, causing this residual chromomagnetic field to vanish. This does not happen, though, because the ghosts and the unphysical gluon do not cancel. This is related to the unitarity problem of the model, which could be solved non-perturbatively in the zero color sector when incorporating confinement. \cite{lcos}

\begin{figure}\begin{center}
\includegraphics[width=5cm]{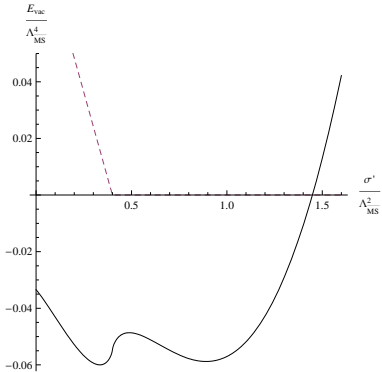}
\caption{Real (full line) and imaginary part (dashed line) of the potential for $H = \unit{0.4}{\Lambda_{\overline{\text{MS}}}^2}$. \label{VeffbijH}}
\end{center}\end{figure}
\begin{figure}\begin{center}
\includegraphics[width=0.48\textwidth]{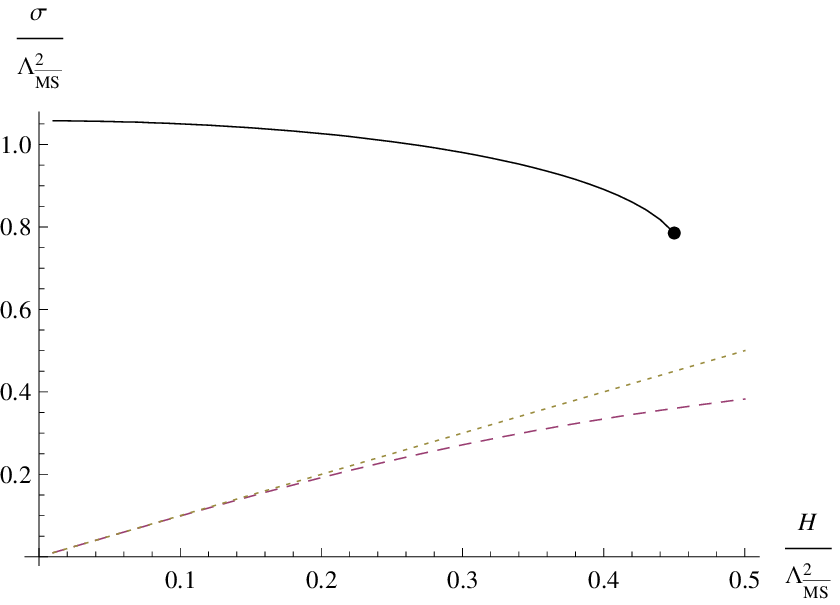} \includegraphics[width=0.48\textwidth]{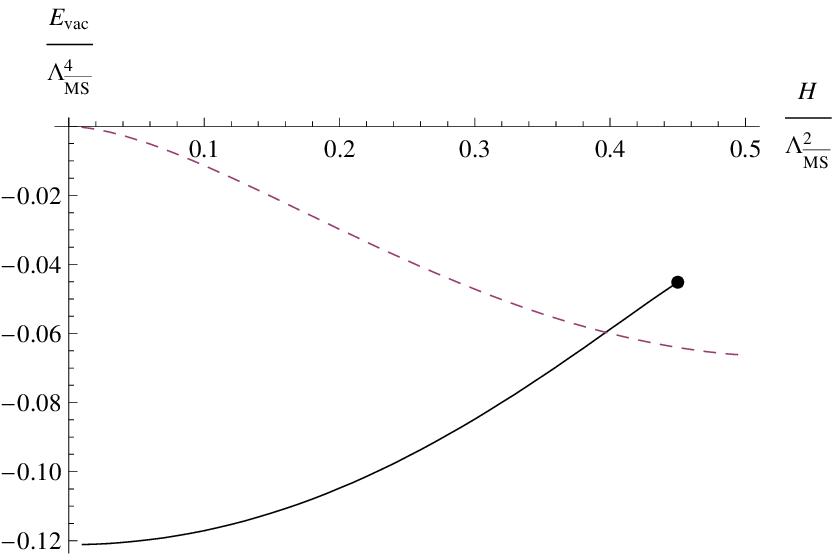}
\caption{\textbf{Left:} The various values of $\sigma'$ as functions of $H$. The full line is the non-perturbative value of $\sigma'$, the dashed line is the value of $\sigma'$ in the lower minimum, and the dotted line is $\sigma'=H$, drawn for reference. \textbf{Right:} The vacuum energy in the minima as a function of $H$. In both plots the thick dot indicates where the higher minimum in the effective potential disappears. \label{functionofH}}
\end{center}\end{figure}

When $H$ is increased, analytic methods have to be abandoned and we solve the equations numerically instead. A qualitative sketch of the effective potential in this regime is depicted in Figure \ref{VeffbijH}. A nonzero imaginary part exists ever when $\sigma'<H$, as mentioned above. In the real part of the action, the point with $\sigma' = 0$ is no longer an extremum, but a new  perturbative minimum forms for $\sigma'$ between zero and $H$. This is separated from the original non-perturbative minimum by a little hill with a top at $\sigma'$ slightly above $H$. The value of $\sigma'$ in the non-perturbative minimum decreases with increasing $H$. For higher $H$ a point is reached where the minimum with smaller $\sigma'$ has a lower energy than the one with greater $\sigma'$. We thus find a first-order phase transition around $H=\unit{0.40}{\Lambda_{\overline{\text{MS}}}^2}$. For $H$ yet higher, the non-perturbative minimum disappears altogether and only the perturbative one remains. These evolutions can be seen in Figure \ref{functionofH}.

The conclusion is that a nonzero chromomagnetic field decreases the effective gluon mass, and when the field is sufficiently high a phase transition occurs, lowering the mass to a value slightly lower than $gH$.

\subsection{Effect of $\sigma'$ on $H$}
We can take the limit $\sigma'=0$, giving
\begin{equation}
\mathcal{L}_\text{eff} (\sigma=0) = \frac12H^2 - \frac{g^2H^2}{(4\pi)^2} \left(4 - \frac{11}3 \ln\frac{gH}{\bar{\mu}^2} + 4\zeta'(-1) + \frac23\ln(2)\right) - \frac{\imath g^2H^2}{8\pi}
\end{equation}
Here we used that $\zeta(s,1/2) = (2^s-1)\zeta(s)$. Ignoring the imaginary part, and putting $\bar\mu^2$ equal to the value of $gH$ in the global minimum, we obtain a perturbative extremum in $H=0$ and a non-perturbative one in
\begin{equation}
gH = \bar{\mu}^2 \exp\left(-\frac{24\pi^2}{11g^2} + \frac{13}{22} + \frac{12}{11} \zeta'(-1) + \frac2{11}\ln(2)\right) \approx \unit{1.71}{\Lambda_{\overline{\text{MS}}}^2}
\end{equation}

When expanding in a series in $\sigma'$, the next term in the real part is
\begin{equation}
-\frac{3g^2H\sigma'}{(4\pi)^2} \ln2
\end{equation}
meaning the non-perturbative minimum will be lowered. This also increases the induced value of $H$ by an amount of $9\sigma'\ln2/22 \approx \unit{0.28}{\sigma'}$.

\begin{figure}\begin{center}
\includegraphics[width=5cm]{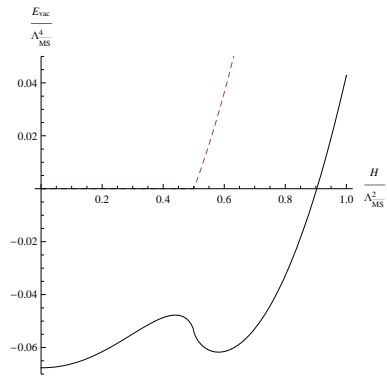}
\caption{Real (full line) and imaginary part (dashed line) of the potential for $\sigma' = \unit{0.5}{\Lambda_{\overline{\text{MS}}}^2}$. \label{Veffbijsigma}}
\end{center}\end{figure}
When going to higher values of $\sigma'$, we find that $H=0$ (or near-to zero) turns into a local minimum of the potential. For $H$ slightly below $\sigma'$ there is a maximum and for $H$ higher than $\sigma'$ there is a non-perturbative minimum (see Figure \ref{Veffbijsigma}). When increasing $\sigma'$, this last one first deepens out, reaching a lowest value for $\sigma'=\unit{0.40}{\Lambda_{\overline{\text{MS}}}^2}$, and it then goes up again. The value of $H$ in this point grows with increasing $\sigma'$. For $\sigma'$ big enough this $H$ asymptotically goes to $\sigma'$.

\begin{figure}\begin{center}
\includegraphics[width=0.48\textwidth]{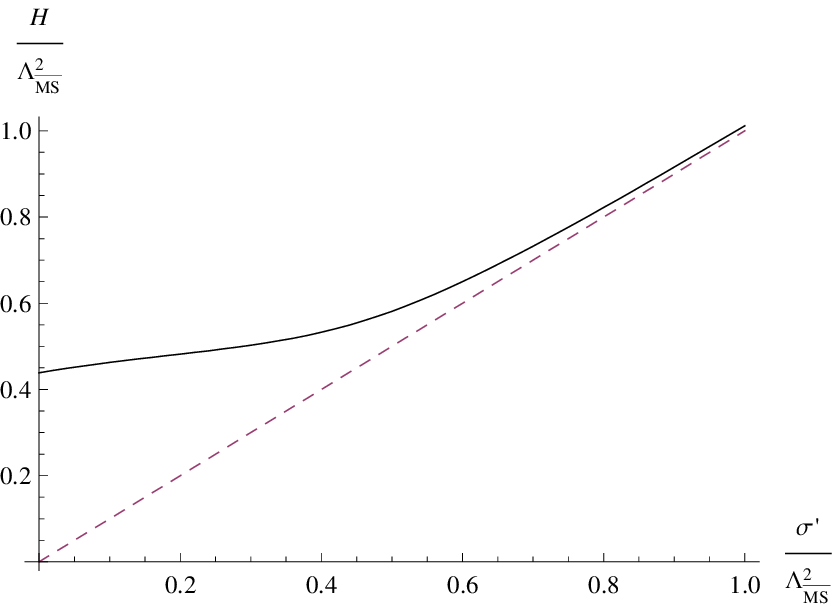} \includegraphics[width=0.48\textwidth]{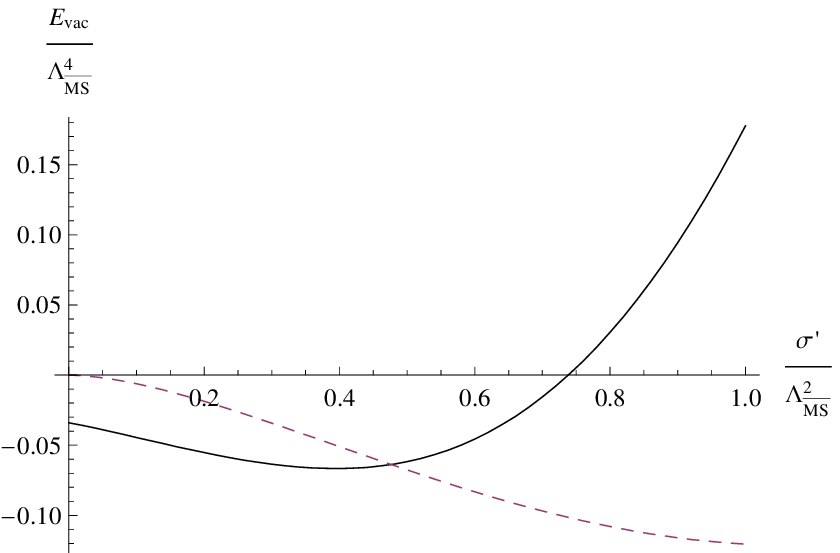}
\caption{\textbf{Left:} The induced value of $H$ as a function of $\sigma'$ (full line). For higher values of $\sigma'$ this nears the asymptotic $H=\sigma'$ (dashed line). \textbf{Right:} The vacuum energies in the non-perturbative minimum (full line) and in $H=0$ (dashed line). The branch for $H=0$ is the same as the potential calculated in \cite{lcos}, reaching its lowest value for $\sigma'_\text{np} = \unit{1.06}{\Lambda_{\overline{\text{MS}}}^2}$. \label{functionofsigma}}
\end{center}\end{figure}
The value of the effective action in $H=0$ decreases for rising $\sigma'$, so that around $\sigma'=\unit{0.48}{\Lambda_{\overline{\text{MS}}}^2}$ it dives lower than the energy in the non-perturbative minimum. This means that, at this point, there is a first-order phase transition from the state with $H>\sigma'$ to the one with $H\approx0$, causing the imaginary part in the action to vanish. This is depicted in Figure \ref{functionofsigma}.

We conclude that switching on a nonzero gluon mass first makes $H$ increase, and then destroys it completely. When the gluon mass is sufficiently large, the vacuum is no longer unstable against the formation of a constant chromomagnetic field, and the Nielsen-Olesen instability, caused by the imaginary part, also is resolved.

\section{Conclusions}\label{conclusions}
We found that, when considering both a constant chromomagnetic field and an $\langle A_\mu^2\rangle$ condensate, the effective action was minimized for zero (or near-to zero) chromomagnetic field with a non-perturbative value for $\langle A_\mu^2\rangle$ as found by Verschelde \emph{et al.} \cite{lcos} There are no unstable modes any longer, and the imaginary part in the action is zero in this minimum.

When considering the situation in which $H$ is an external field, we found that applying such a field first lowers the value of the induced mass, and for $H$ around $\unit{0.40}{\Lambda_{\overline{\text{MS}}}^2}$ the non-perturbative mass is destroyed, leaving only a perturbative value slightly smaller than $H$. The action then has a (small) imaginary part as in the Savvidy case.

When, on the other hand, considering the effect of the mass on the Savvidy field, we found that a sufficiently high gluon mass ($\sigma'\geq\unit{0.48}{\Lambda_{\overline{\text{MS}}}^2}$) destroys the induced $H$ field, at the same time causing the Nielsen-Olesen instability (the imaginary part in the action) to vanish.

\section*{Acknowledgements}
One of us (DV) would like to thank David Dudal for helpful discussions. This work is supported financially by the "Special Research Fund" of Ghent University.

\bibliographystyle{unsrt}

\end{document}